# High temperature collective spin-photon coupling in a microwave cavity

E.I. Baibekov

*Kazan Federal University, 420008 Kazan, Russian Federation*
*e-mail: edbaibek@gmail.com*

**Abstract**

An ensemble of $N$ identical noninteracting spins being in thermal equilibrium and coupled to the resonant mode of a lossless microwave cavity is studied at arbitrary temperature $\tau$. Near $\tau = 0$ the system is known to be in a coupled spin-photon state that manifests itself by the splitting of the cavity mode (vacuum Rabi splitting). The cavity emission spectrum is simulated for arbitrary $\tau$. It is shown that the spin-photon coherence can be partially preserved for $\tau < \omega_S \sqrt{N}/2$, where $\omega_S$ is the spin excitation energy, even in case when the spins are randomly directed. The calculations corroborate recent room-temperature observations of the collective coupling between the microwave cavity mode and the electron spin ensemble (NV centers in diamond, DPPH, $Fe_8$ nanomagnets). At higher $\tau$, as a consequence of thermal excitations within the spin ensemble, the two lines of the emission spectrum merge into a narrow line with broad wings.

## I. Introduction

Quantum coherence is a fundamental concept of modern physics that reveals itself in wide range of phenomena either of matter or light origin. Being naturally intrinsic to microscopic systems [1], it has applications on a macroscopic level, starting from the canonical examples (lasers, superconductivity, superfluidity), to the modern advances in quantum information processing [2], quantum cryptography [3] and quantum teleportation [4]. In cavity quantum electrodynamics, photons in a reflective cavity are coupled to atoms or spins [5,6] thus providing an opportunity to combine and observe both light and matter coherence. Among the successful implementations are Rydberg atoms [7,8], trapped ions [9] and semiconductor quantum dots [10], all of them coupled to the photonic field via electric dipole transitions. Recently, several experimental groups reported the observations of magnetic-type coupling that involved interactions of light with the particles bearing electron spin [11-17]. While the electric dipole transitions provide stronger coupling, the spin-based systems usually maintain longer coherence times [18]. The weaker magnetic coupling means either the necessity of much more precise detection or the substantial increase in the number of spins in a cavity. The examples of the latter are a powder of $10^{18}$ DPPH radicals [14], the ensembles of $10^{16}$ $Fe_8$ molecular nanomagnets [16] and $10^{11}$ nitrogen-vacancy centers in diamond [17]. The coupling with the cavity mode can be achieved with standard electron paramagnetic resonance (EPR) instrumentation [13,14]. In all these experiments, well-resolved splitting of the cavity mode was an indication of the strong coupling regime based on the coherence of the spin and photon states. Since each cavity photon interacts simultaneously with all the spins in the cavity, there is an opportunity to transfer information coherently from the photons to the spin ensemble as a whole, with the applications in quantum computing [19,20] and quantum holography [21].

The splitting of the cavity mode in the strong coupling regime is analogous to the well-known vacuum-field Rabi splitting predicted for a single excited two-level atom in the absence of the cavity field (or, equivalently, for an unexcited atom and a single cavity photon) [22]. The same is found for an ensemble of $N$ unexcited two-level atoms, or spins in the external magnetic field, interacting with a single photon in the cavity [23]. Further we will focus on the ensemble of $N \gg 1$ spins (paramagnetic ions, crystal defects, single molecule magnets, etc.) incorporated into the solid or liquid medium and being in thermal equilibrium with the host material. The above-mentioned case would correspond to zero temperature of the spin ensemble, $\tau = 0$. The problem has an approximate solution if the numbers of excited spins and cavity photons are both $\ll N$, i.e. when $\tau$ is close to zero and the spin ensemble is still highly polarized [24-26]. The calculation of the cavity energy spectrum in the general case of arbitrary $\tau$ represents a challenging task. A typical X-band microwave cavity has the lowest-order mode frequency $\omega_0/2\pi \sim 10$ GHz. Under resonant conditions that would correspond to the spin excitation energy $\sim 0.5$ K. Such spin ensemble taken at room temperature would be in a disordered state, with nearly half of the spins excited. However, as we will show, even in this case the spin-photon coherence is partially maintained.

The paper is organized as follows. Section 2 is a brief revision of Tavis-Cummings Hamiltonian [27] and its lowest energy solutions that correspond to the vacuum Rabi splitting. The calculations of the cavity emission spectrum at $\tau \ll \omega_S \sqrt{N}/2$ are presented in Section 3. Section 4 describes the phenomenon of thermal decoherence related to the spin polarization disorder in the case when the spin ensemble is subjected to extreme heating.

**II. Tavis-Cummings Hamiltonian**

Let us consider an ensemble of $N \gg 1$ identical noninteracting spins-1/2 coupled with the single-mode radiation field of a lossless microwave cavity. We adopt Tavis-Cummings model [27] and write the system Hamiltonian in the rotating wave approximation as

$$\hat{H} = \omega_0 \hat{a}^+ \hat{a} + \omega_S \hat{S}_z + g\left(\hat{a}\hat{S}_+ + \hat{a}^+ \hat{S}_-\right), \qquad (1)$$

where $\omega_0$ is the photon frequency, $\omega_S$ is the spin excitation energy, $\hat{a}^+$ and $\hat{a}$ are the photonic creation and annihilation operators. The last term in Eq. (1) represents the spin-photon interaction. On condition that the host sample dimensions are much smaller than the field wavelength, the coupling constant $g$ is the same for all spins in the sample. Let us denote by $|n_0\rangle$ the state of the cavity containing $n_0$ photons, $\hat{a}^+ \hat{a} |n_0\rangle = n_0 |n_0\rangle$. Since the Hamiltonian is written in terms of collective spin operators $\hat{S}_z = \sum_{j=1}^{N} \hat{s}_{jz}$ and $\hat{S}_\pm = \sum_{j=1}^{N} \hat{s}_{j\pm}$, it is convenient to introduce Dicke states $|(\gamma S) S_z\rangle$ [28]. The collective spin $S$ (sometimes called the cooperation number) and the spin projection $S_z$ satisfy the relations

$$\hat{S}^2 |(\gamma S) S_z\rangle = S(S+1) |(\gamma S) S_z\rangle, \quad \hat{S}_z |(\gamma S) S_z\rangle = S_z |(\gamma S) S_z\rangle. \qquad (2)$$

The index $\gamma$ enumerates different sets of spin states with the same $S$, and $|S_z| \leq S \leq N/2$. The Hamiltonian (1) commutes with $\hat{S}^2$ and with $\hat{a}^+\hat{a} + \hat{S}_z$, the last being related to the total number of excitations $n = n_0 + S_z + S$ in the system. A set of eigenstates of (1) corresponding to a certain set $\gamma S$ and excitation number $n$ may be expressed as

$$|(\gamma S n)\alpha\rangle = \sum_{n_0 + S_z + S = n} C_{n_0 S_z \alpha}^{(\gamma S n)} |n_0\rangle |(\gamma S) S_z\rangle, \qquad (3)$$

with index $\alpha = 1, 2, \ldots, \min\{n_0+1, 2S+1\}$ enumerating all possible eigenstates related to the same $\gamma, S, n$. For example, the ground state of the spin ensemble has $S_z = -N/2$ (all spins are down) and belongs to the ground set $S = N/2$. Since this is the only $S = N/2$ set of the spin system, we can omit $\gamma$ in this case. The spin-photon state $|1\rangle|N/2, -N/2\rangle$ corresponding to $n=1$ is mixed only with the vacuum-field state $|0\rangle|N/2, -N/2+1\rangle$. At resonance, $\omega_S = \omega_0$, these two energy levels are split by the gap of $2g\sqrt{N}$ [27]. While the expansion (3) is generally valid, the calculation of the coefficients $C_{n_0 S_z \alpha}^{(\gamma S n)}$ and the energy levels for arbitrary $n$ represents an intractable problem. It is possible, however, to obtain approximate solutions in the case $n \ll S$ introducing bosonic operators $\hat{b}$ and $\hat{b}^+$ by means of Holstein-Primakoff transformation [24,26]

$$\hat{S}_z = -S + \hat{b}^+\hat{b}, \quad \hat{S}_- = \sqrt{2S - \hat{b}^+\hat{b}} \cdot \hat{b} \approx \sqrt{2S}\hat{b}, \quad \hat{S}_+ = \hat{b}^+\sqrt{2S - \hat{b}^+\hat{b}} \approx \sqrt{2S}\hat{b}^+. \qquad (4)$$

Within the set $\gamma S$, the Hamiltonian (1) becomes quadratic in the bosonic operators $\hat{a}, \hat{a}^+, \hat{b}, \hat{b}^+$

$$\hat{H} = \omega_0 \hat{a}^+\hat{a} + \omega_S \hat{b}^+\hat{b} + \Omega\left(\hat{a}\hat{b}^+ + \hat{a}^+\hat{b}\right) - S\omega_S, \qquad (5)$$

where $\Omega = g\sqrt{2S}$ is Rabi frequency that corresponds to the cooperation number $S$. It is diagonalized by means of a linear transformation $\hat{a} = \kappa_1 \hat{c} + \kappa_2 \hat{d}$, $\hat{b} = \kappa_1 \hat{d} - \kappa_2 \hat{c}$, with

$$\kappa_{1,2} = \frac{1}{\sqrt{2}}\sqrt{1 \mp \frac{\Delta}{\sqrt{\Delta^2 + 4\Omega^2}}}, \quad \Delta = \omega_S - \omega_0, \quad \kappa_1^2 + \kappa_2^2 = 1, \qquad (6)$$

so that

$$\hat{H} = \omega_c \hat{c}^+\hat{c} + \omega_d \hat{d}^+\hat{d} - S\omega_S, \quad \omega_{c,d} = \omega_0 + \frac{\Delta}{2} \pm \frac{1}{2}\sqrt{\Delta^2 + 4\Omega^2}. \qquad (7)$$

The eigenstates (3) and their energies are now defined by the occupation numbers $n_c$ and $n_d$ of the polariton modes $\hat{c}$ and $\hat{d}$, with $n_c + n_d = n$:

$$\left|(\gamma Sn)\alpha\right\rangle \equiv \left|(\gamma S)n_c n_d\right\rangle, \quad E_{Sn_c n_d} = \omega_c n_c + \omega_d n_d - S\omega_S. \qquad (8)$$

At resonance $(\Delta = 0)$, the spin-photon coupling corresponds to the highest possible mixing of the bosonic modes $\hat{a}$ and $\hat{b}$: $\hat{c} = \frac{\hat{a} - \hat{b}}{\sqrt{2}}$, $\hat{d} = \frac{\hat{a} + \hat{b}}{\sqrt{2}}$. Conversely, far from resonance $(|\Delta| \gg \Omega)$, the spin-photon coupling is negligible, and $\hat{c}, \hat{d} \to \hat{a}, \hat{b}$.

## III. Cavity emission spectrum at low and medium temperatures

Let us assume that both the Rabi frequency and the detuning are small compared to the cavity frequency and the temperature:

$$\Omega, |\Delta| \ll \omega_0, \tau. \qquad (9)$$

These conditions are generally valid in EPR and are required to obtain a coupled state. We calculate the emission spectrum of the cavity as

$$G(\omega) = \frac{\sum_{\Psi, \Phi} |a_{\Phi\Psi}|^2 \rho_{\Psi\Psi} \delta(E_\Psi - E_\Phi - \omega)}{\sum_{\Psi, \Phi} |a_{\Phi\Psi}|^2 \rho_{\Psi\Psi}}, \qquad (10)$$

where $|\Psi\rangle = |(\gamma Sn)\alpha\rangle$ is the initial collective state with energy $E_\Psi$, $|\Phi\rangle = |(\gamma S, n-1)\beta\rangle$ is the final state resulting from the emission of a single photon with frequency $\omega$, $a_{\Phi\Psi} = \langle\Phi|\hat{a}|\Psi\rangle$, $\hat{\rho}$ is

the collective density matrix, $\rho_{\Psi\Psi} \sim \exp(-E_\Psi/\tau)$ is the probability to occupy the state $|\Psi\rangle$, and $G(\omega)$ is normalized to unity. As was first indicated by Dicke [28], for a macroscopic spin ensemble in thermal equilibrium, both $S_z$ and $S$ are well-defined and satisfy the relation $S_z \simeq \langle S_z \rangle \simeq -S$, where $\langle ... \rangle$ denotes thermal averaging over the spin states,

$$\langle S_z \rangle = -\frac{N}{2}\tanh\frac{\omega_S}{2\tau}. \tag{11}$$

More precisely, the spin temperature must meet the condition $\tau \ll \omega_S \sqrt{N}/2$, or, equivalently, $\sqrt{N} \ll 2|\langle S_z \rangle| \leq N$ (see Appendix). Up to room temperature, this is usually fulfilled for $N > 10^6$. Supposing that the cavity field is weak $(n_0 \ll S)$, so it is unable to alter sufficiently the thermal equilibrium state of the spin ensemble, we can use the approximate solutions (8), where $|\Psi\rangle = |(\gamma S)n_c n_d\rangle$, and $|\Phi\rangle$ equals either $|(\gamma S)n_c - 1, n_d\rangle$ or $|(\gamma S)n_c, n_d - 1\rangle$:

$$G(\omega) = \frac{\sum_{Sn_c n_d} \eta(S)\{\kappa_1^2 n_c \delta(\omega-\omega_c) + \kappa_2^2 n_d \delta(\omega-\omega_d)\}\exp(-E_{Sn_c n_d}/\tau)}{\sum_{Sn_c n_d} \eta(S)\{\kappa_1^2 n_c + \kappa_2^2 n_d\}\exp(-E_{Sn_c n_d}/\tau)}. \tag{12}$$

Summation over $\gamma$ gives the number of sets with the same $S$ denoted above by $\eta(S)$. For $S \simeq |\langle S_z \rangle|$ it is approximated by (see Appendix)

$$\eta(S) \simeq \eta(|\langle S_z \rangle|)\exp\left\{-\frac{2N(S-|\langle S_z \rangle|)^2}{N^2 - 4|\langle S_z \rangle|^2} - \frac{\omega_S(S-|\langle S_z \rangle|)}{\tau}\right\}. \tag{13}$$

Since $S \gg 1$, one can use the continuum approximation $\sum_S ... \to \int dS ...$ The integration over $S$ and summation over $n_c, n_d$ finally yield

$$G(\omega) = \frac{1}{\sqrt{2\pi}\sigma}\left\{\kappa_1^2 \exp\left(-\frac{(\omega-\omega_c)^2}{2\sigma^2}\right) + \kappa_2^2 \exp\left(-\frac{(\omega-\omega_d)^2}{2\sigma^2}\right)\right\},$$
$$\sigma = \frac{\Omega^2}{\sqrt{\Delta^2 + 4\Omega^2}}\sqrt{\frac{N}{4\langle S_z \rangle^2} - \frac{1}{N}}. \tag{14}$$

The quantities $\kappa_{1,2}$, $\omega_{c,d}$ and $\Omega$ as functions of $S$ are now averaged over temperature, with $S = |\langle S_z \rangle|$. The emission spectrum consists of two Gaussian lines centered at $\omega_c$ and $\omega_d$. They are separated by the gap $\omega_c - \omega_d = \sqrt{\Delta^2 + 4\Omega^2}$ and have equal standard deviations $\sigma$. The average Rabi frequency $\Omega = g\sqrt{2|\langle S_z \rangle|}$, $\omega_{c,d}$ and $\sigma$ are temperature-dependent.

Let us analyze the behavior of $G(\omega)$ in different temperature intervals. At low temperatures, when $0 \le \tau \ll \omega_S/2$, the spin ensemble is close to its ground state with $\langle S_z \rangle = -N/2$ (see Eq. (11)). The emission spectrum of the spin-photon system degenerates into two Dirac delta functions

$$G(\omega) = \kappa_1^2 \delta(\omega - \omega_c) + \kappa_2^2 \delta(\omega - \omega_d), \qquad (15)$$

with the gap of $\sqrt{\Delta^2 + 4g^2 N}$. The results in this limiting case are consistent with the lowest-energy solutions of Tavis-Cummings Hamiltonian [23] and, if resonance condition $\Delta = 0$ is fulfilled, give vacuum Rabi splitting $2g\sqrt{N}$ with the highest possible spin-photon coupling. $G(\omega)$ does not vary with temperature as long as $0 \le \tau \ll \omega_S/2$. As $\tau$ becomes comparable to $\omega_S$, the average cooperation number decreases in accordance with Eq. (11), and so does the Rabi splitting. In the region $1 \ll 2\tau/\omega_S \ll \sqrt{N}$ that we will further call "medium temperatures", $|\langle S_z \rangle| \simeq N\omega_S/4\tau \ll N/2$, so that the numbers of spin-up and spin-down states in the spin ensemble become almost equal. The two delta functions broaden into Gaussian lines with $\sigma = g\sqrt{\tau/2\omega_S}$, while the Rabi splitting $2\Omega$ which determines the collective coupling strength decreases to $g\sqrt{2N\omega_S/\tau}$ (the resonance case $\omega_S = \omega_0$ is assumed here for simplicity). The collective spin-photon coupling is still possible in this temperature interval as soon as the two lines are well resolved. Though, in order to reach the same coupling strength as at lower temperatures, one needs to increase either the number of spins $N$ in the cavity or the coupling constant $g$. Typical experimental conditions (X-band 10 GHz microwave cavity) enable successful room-temperature coupling for $N > 10^6$ (see the simulations of the cavity spectrum in Fig. 1). The results scale as $\sqrt{N}\omega_S\tau^{-1}$: e.g., the distribution $G(\omega)$ calculated in Fig. 1a for $N = 10^9$ and $\tau = 300$ K remains the same in the model parameters for $N = 10^5$ and $\tau = 3$ K. The last distribution obtained for $N = 10^6$ spins (Fig. 1d) lies outside the medium temperature region and is therefore only a rough approximation.

The calculated positions and relative heights of the two lines as functions of detuning and $\langle S_z \rangle$ agree with the results obtained using a simple model of two coupled oscillators [14,17]. Particularly interesting, however, are the lineshapes and their half-widths since they affect the lifetimes of the spin-photon state. The model of two coupled oscillators predicts Lorentzian lineshapes with the half-width depending on the cavity quality factor and the spin decay rate, both being phenomenological parameters. Our direct calculations result in Gaussian distribution, while the corresponding half-width is given explicitly (14). Note that we neglect cavity losses and the relaxation within the spin ensemble. The obtained broadening is attributed solely to thermal disorder of the collective spin state.

**IV. High temperatures: thermal decoherence inside the spin-photon ensemble**

Thermal excitations inside the spin ensemble that arise as a result of subsequent heating would eventually destroy the coherent spin-photon state. The two lines of the emission spectrum merge into one, and the Rabi splitting becomes unresolved. As follows from the previous section, the critical temperature of this process is $\tau_C = \omega_S \sqrt{N}/2$. Note that $\tau_C$ depends on the number of spins: assuming the same 10 GHz cavity, $\tau_C = 240$ K for $N = 10^6$, 2.4 K for $N = 100$, etc. Direct calculation of $G(\omega)$ in the temperature range $\tau \geq \tau_C$ seems practically impossible since it requires complete solution for the Tavis-Cummings Hamiltonian, and not only its lower-energy levels. The excitation number $n$ is no longer $\ll S$, so one cannot replace the spin excitations with the bosonic modes (4), $S$ and $S_z$ are no longer well-defined, and many of the other approximations do not work in this case [29]. Not attempting to solve this, possibly, insoluble task, let us construct $G(\omega)$ indirectly. Under certain condition, one can calculate its first several moments and estimate the rest. For simplicity, we restrict ourselves to the resonance case. The *k*th central moment of $G(\omega)$ is defined as

$$m_k = \int \omega^k G(\omega + \omega_S) d\omega = \frac{\sum_{\Psi,\Phi} |a_{\Phi\Psi}|^2 \rho_{\Psi\Psi} (V_{\Psi\Psi} - V_{\Phi\Phi})^k}{\sum_{\Psi,\Phi} |a_{\Phi\Psi}|^2 \rho_{\Psi\Psi}}, \qquad (16)$$

where we use the expression (10), and $\hat{V} = g(\hat{a}\hat{S}_+ + \hat{a}^+\hat{S}_-)$ is the interaction part of the Hamiltonian (1). A specific rearrangement $\langle \Psi | \hat{a}^+ | \Phi \rangle (V_{\Psi\Psi} - V_{\Phi\Phi}) = \langle \Psi | [\hat{V}, \hat{a}^+] | \Phi \rangle$, performed *k* times, gives

$$m_k = \frac{\mathrm{Tr}\left(\overbrace{[\hat{V},[\hat{V},\ldots[\hat{V},\hat{a}^+]]]}^{k\text{ times}}\hat{a}\hat{\rho}\right)}{\mathrm{Tr}(\hat{a}^+\hat{a}\hat{\rho})}. \qquad (17)$$

The density matrix of the system in thermal equilibrium is

$$\hat{\rho} = \frac{\exp\{-(\hat{H}_0 + \hat{V})/\tau\}}{\mathrm{Tr}\exp\{-(\hat{H}_0 + \hat{V})/\tau\}}, \qquad (18)$$

where $\hat{H}_0 = \omega_S(\hat{a}^+\hat{a} + \hat{S}_z)$ is the zero-order Hamiltonian. Assuming that the coupling $g$ is small enough, we can neglect $\hat{V}$ in (18) and separate the spin and photonic operators:

$$\hat{\rho} \simeq \frac{\exp\{-\omega_S \hat{S}_z/\tau\}}{\text{Tr}_S \exp\{-\omega_S \hat{S}_z/\tau\}} \cdot \frac{\exp\{-\omega_S \hat{a}^+ \hat{a}/\tau\}}{\text{Tr}_{Ph} \exp\{-\omega_S \hat{a}^+ \hat{a}/\tau\}} = \hat{\rho}_S \cdot \hat{\rho}_{Ph}. \tag{19}$$

The operator $\hat{\rho}_S$ coincides with the density matrix of thermally equilibrated free spin ensemble. Under this approximation, the calculation of traces in (17) is straightforward for the first few $k$, giving

$$m_2 = \Omega^2, \quad m_4 = \Omega^4 \left(1 + \frac{1}{N}\coth^2 \frac{\omega_S}{2\tau}\right), \quad m_6 = \Omega^6 \left(1 + \frac{3}{N}\coth^2 \frac{\omega_S}{2\tau} + \frac{4}{N^2}\coth^4 \frac{\omega_S}{2\tau}\right). \tag{20}$$

Using the commutation relations for the spin and bosonic operators, one can derive the general form of $k$th moment ($C_{kj}$ are numeric constants):

$$m_{2k-1} = 0, \quad m_{2k} = \Omega^{2k} \left\{1 + \sum_{j=1}^{k-1} \frac{C_{kj}}{N^j} \coth^{2j} \frac{\omega_S}{2\tau}\right\}. \tag{21}$$

Since only the even order moments are non-zero, the line shape at resonance is symmetric. At low temperatures, $m_{2k} \simeq \Omega^{2k}$, meaning that $G(\omega)$ is well represented by two delta functions centered at $\omega \pm \Omega$, in full agreement with Section III. The temperature evolution of the emission spectrum can be traced by the 4$^{\text{th}}$ standardized moment, $m_4/m_2^2$. In the medium temperature range, it grows very slowly with $\tau$ and is still close to unity (the left half of Fig. 2a and the right half of Fig. 2b). It doubles at the critical point $\tau = \tau_C$ (the dashed line in Fig. 2) and grows quadratically with $\tau$ after that. The higher-order moments show similar temperature behavior. In the high temperature interval defined as $1 \ll \tau/\tau_C \ll N^{1/2}$, we obtain

$$m_2 = \Omega^2, \quad \frac{m_4}{\Omega^4} \simeq \left(\frac{\tau}{\tau_C}\right)^2, \quad \frac{m_6}{\Omega^6} \simeq 4\left(\frac{\tau}{\tau_C}\right)^4, \ldots, \quad \frac{m_{2k}}{\Omega^{2k}} \sim \left(\frac{\tau}{\tau_C}\right)^{2k-2}. \tag{22}$$

It is peculiar that the standard deviation of the distribution, $\sqrt{m_2}$, which is related to its half-width, equals Rabi frequency in the whole temperature range. At the same time, the growth of higher-order moments indicates that $G(\omega)$ has long wings at high $\tau$. For a given infinite set of $m_{2k}$, one can obtain the characteristic function of the distribution

$$\chi(t) = \int e^{i\omega t} G(\omega + \omega_S) d\omega = \sum_{k=0}^{\infty} \frac{(-1)^k m_{2k} t^{2k}}{(2k)!}. \tag{23}$$

A Fourier transformation of (23) within the approximation (22) gives the general form of $G(\omega)$

$$G(\omega) = \left[1 - \theta\left(\frac{\tau}{\tau_C}\right)^{-2}\right] g_0(\omega) + \theta\left(\frac{\tau}{\tau_C}\right)^{-2} g_1(\omega), \tag{24}$$

which is a superposition of two symmetric distributions, $g_0(\omega)$ and $g_1(\omega)$, both centered at $\omega = \omega_S$ and normalized to unity; $\theta$ is a numeric constant of the order unity. The two components have different weights and widths. In particular, $g_0(\omega)$ is tall but narrow, with the half-width $\sim \Omega$, while $g_1(\omega)$ is broad (the half-width $\sim \Omega\tau/\tau_C$) but low. Assuming that both lines are Gaussian with the standard deviations $\sigma_0$ and $\sigma_1$, respectively, and comparing (24) with (20), we obtain

$$\sigma_0 = 0.76\Omega, \quad \sigma_1 = 0.89\Omega\tau/\tau_C, \quad \theta = 0.52. \tag{25}$$

Fig. 3 shows a simulation of $G(\omega)$ for a system of $N = 10^4$ spins ($\tau_C = 24$ K), thermally equilibrated at $\tau = 100$ K. The Rabi splitting is now unresolved since thermal excitations destroy the coherent spin-photon state. We can call this process "thermal decoherence", and it is related to polarization disorder within the spin ensemble at high temperatures. For even higher temperatures, when $\tau/\tau_C > \sqrt{N}$, the number of photons in the cavity becomes larger than $N$, meaning that the photon field can be treated classically. The emission spectrum contains only the cavity mode line. The results obtained at certain temperature intervals, showing qualitatively different energy spectra, are summarized in Table 1.

**Appendix. Calculation of the cooperation number distribution**

*a) The correlation of $S$ and $\langle S_z \rangle$*

For the sake of clarity, let us reproduce the argumentation presented in [28]. Suppose that the spin ensemble is close to its thermal equilibrium state. The average value $\langle S_z \rangle$ is then given by Eq. (11), and

$$\langle S_z^2 \rangle = \frac{1}{4}\sum_{j,k=1}^{N}\langle \sigma_{jz}\sigma_{kz}\rangle = \frac{1}{4}\sum_{j=1}^{N}\langle \sigma_{jz}^2\rangle + \frac{1}{4}\sum_{\substack{j,k=1 \\ (j\neq k)}}^{N}\langle \sigma_{jz}\rangle\langle \sigma_{kz}\rangle = \frac{N}{4} + \frac{N-1}{N}\langle S_z\rangle^2, \tag{26}$$

where $\hat\sigma_{jz}$ is Pauli matrix of the spin $j$, with $\hat\sigma_{jz}^2 = 1$. The standard deviation of $S_z$ from its average value equals

$$\sigma(S_z) = \sqrt{\frac{N}{4} - \frac{\langle S_z\rangle^2}{N}} \leq \frac{\sqrt{N}}{2}. \tag{27}$$

It follows that if $\sqrt{N} \ll 2\langle S_z\rangle \leq N$, or, equivalently, $\tau \ll \omega_S\sqrt{N}/2$, the quantum number $S_z$ is very close to $\langle S_z\rangle$, or "well-defined". Analogous calculations applied for the cooperation number $S$, assuming that $S_z = \langle S_z\rangle$, give

$$\langle \mathbf{S}^2 \rangle = \langle S_z \rangle^2 + \frac{N}{2}, \quad \sigma(\mathbf{S}^2) = \sqrt{\frac{N^2}{4} - \langle S_z \rangle^2} \leq \frac{N}{2}. \tag{28}$$

Again, if the above-mentioned condition is satisfied, $S$ is well-defined: $\langle \mathbf{S}^2 \rangle \simeq \langle S_z \rangle^2$, $S \simeq |\langle S_z \rangle|$.

*b) Cooperation number distribution near $S = |\langle S_z \rangle|$*

As a matter of fact, there are numerous sets of spin states that correspond to the same cooperation number $S$. If we denote the number of such sets by $\eta(S)$, then

$$\eta(S) = C\left(N, \frac{N}{2} - S\right) - C\left(N, \frac{N}{2} - S - 1\right), \tag{29}$$

where the binomial coefficient $C\left(N, \frac{N}{2} - M\right)$ is the number of collective spin states for which $\frac{N}{2} - M$ spins are up and $\frac{N}{2} + M$ are down, giving a total $S_z = -M$. Since generally $|S_z| \leq S$, then $C\left(N, \frac{N}{2} - M\right)$ is the total number of sets with $S \geq M$, and the difference (29) equals $\eta(S)$. On condition that $N, S \gg 1$, one can use Stirling's approximation for large factorials, expand the expression $\ln \eta(S)$ into Tailor series over the small parameter $(S - |\langle S_z \rangle|)/(N \pm 2|\langle S_z \rangle|)$ and finally arrive to

$$\eta(S) \simeq \eta(|\langle S_z \rangle|) \exp\left\{-\frac{2N(S - |\langle S_z \rangle|)^2}{N^2 - 4|\langle S_z \rangle|^2} - (S - |\langle S_z \rangle|) \ln \frac{N + 2|\langle S_z \rangle|}{N - 2|\langle S_z \rangle|}\right\}, \tag{30}$$

where the terms in the expansion of the order higher than $(S - |\langle S_z \rangle|)^2$ were neglected. Since $\ln \frac{N + 2|\langle S_z \rangle|}{N - 2|\langle S_z \rangle|} = \frac{\omega_S}{\tau}$, one obtains the Gaussian distribution (13).

**Acknowledgments**

This work was supported by RFBR (grant no. 12-02-31336) and by Dynasty Foundation. Author thanks B. Z. Malkin for advice and B. Barbara for discussions which initiated this work.

**Figure captions:**

Fig. 1. Density plot showing the emission spectrum $G(\omega)$ calculated according to Eq. (14) as the function of detuning $\omega_S - \omega_0$ and of frequency $\omega$, both expressed in units of Rabi frequency $\Omega$. The $N$-spin ensemble is positioned inside the $X$-band cavity ($\omega_0/2\pi = 10$ GHz) and is thermally equilibrated at $\tau = 300$ K. The insets to the right of each Fig. show cross-section of the distribution at resonance $(\omega_S = \omega_0)$. (a) $N = 10^9$. (b) $N = 10^8$. (c) $N = 10^7$. (d) $N = 10^6$.

Fig. 2. Standardized central moments $m_{2k}/\Omega^{2k}$ calculated according to Eq. (20) ($\omega_S/2\pi = \omega_0/2\pi = 10$ GHz). The dashed line is related to the critical temperature $\tau_C = \omega_S \sqrt{N}/2$. (a) $N = 10^4$. (b) $\tau = 300$ K.

Fig. 3. A simulation of $G(\omega)$ (thick solid line) by two Gaussians: $g_0(\omega)$ (dashed) and $g_1(\omega)$ (thin solid) for a system of $N = 10^4$ spins, $\omega_S/2\pi = \omega_0/2\pi = 10$ GHz, $\tau = 100$ K, $\tau_C = 24$ K.

Fig. 1a

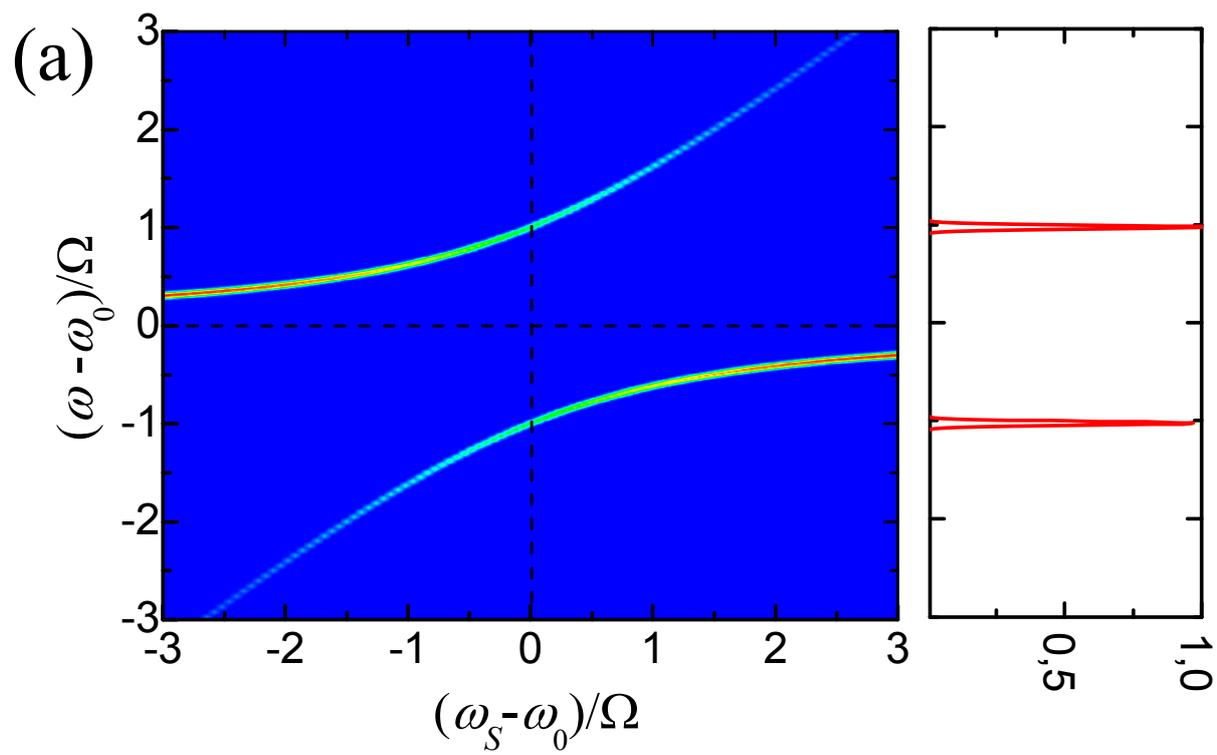

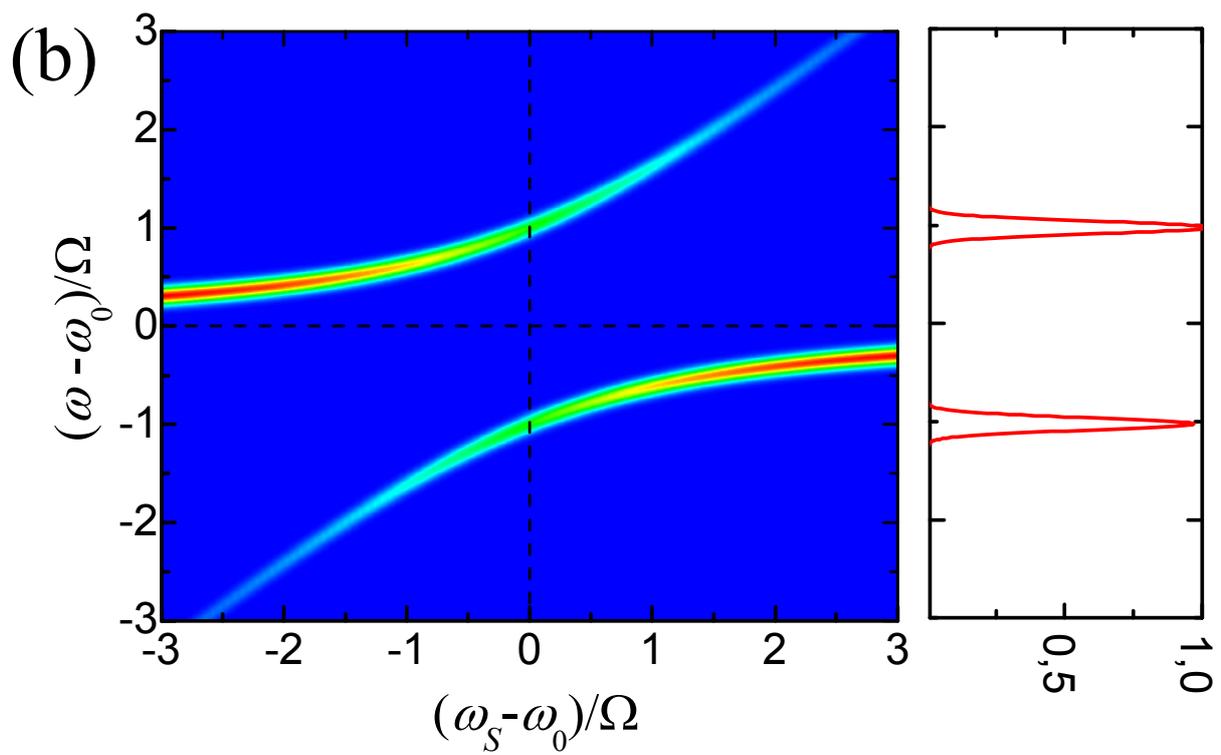

Fig. 1c

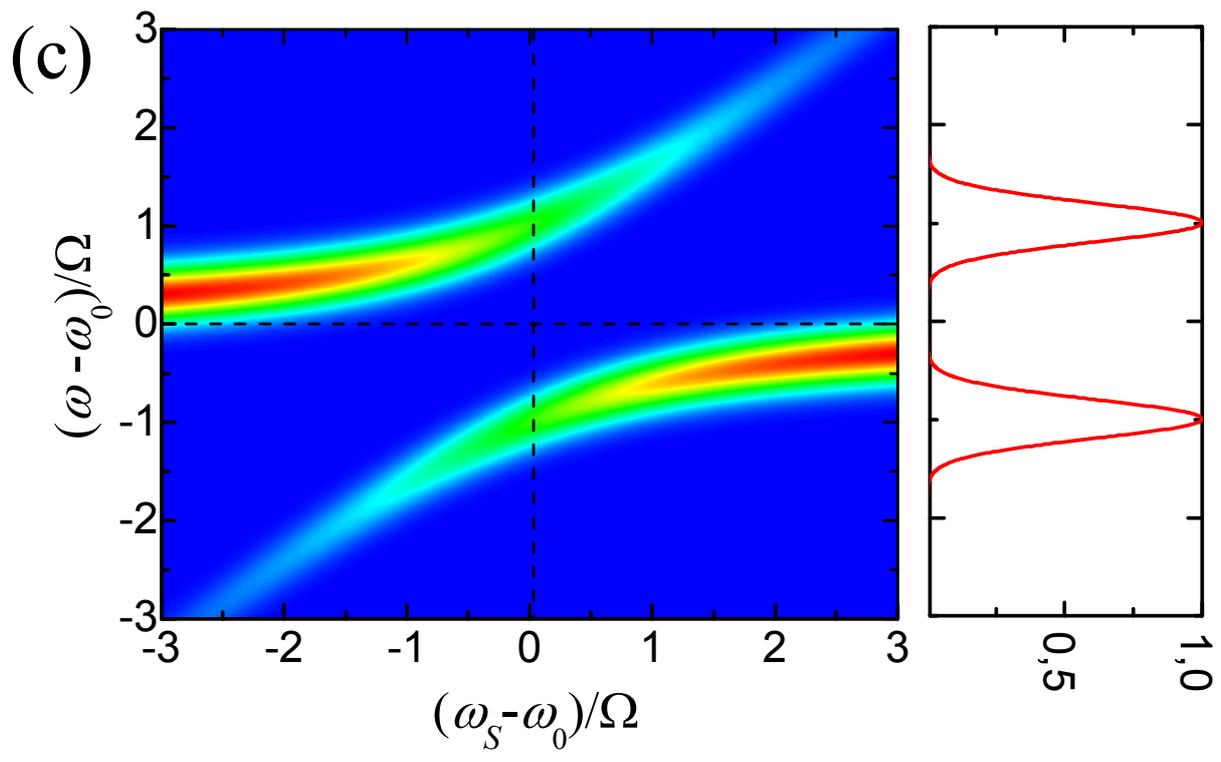

Fig. 1d

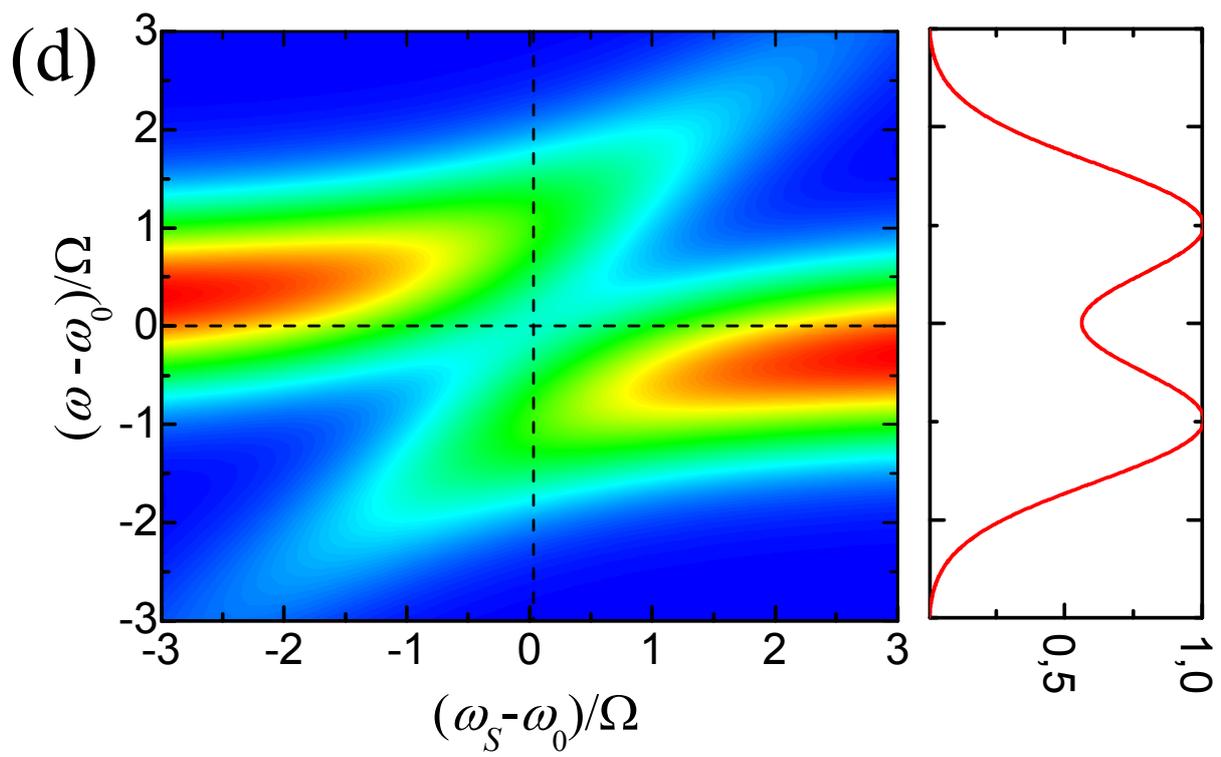

Fig. 2a

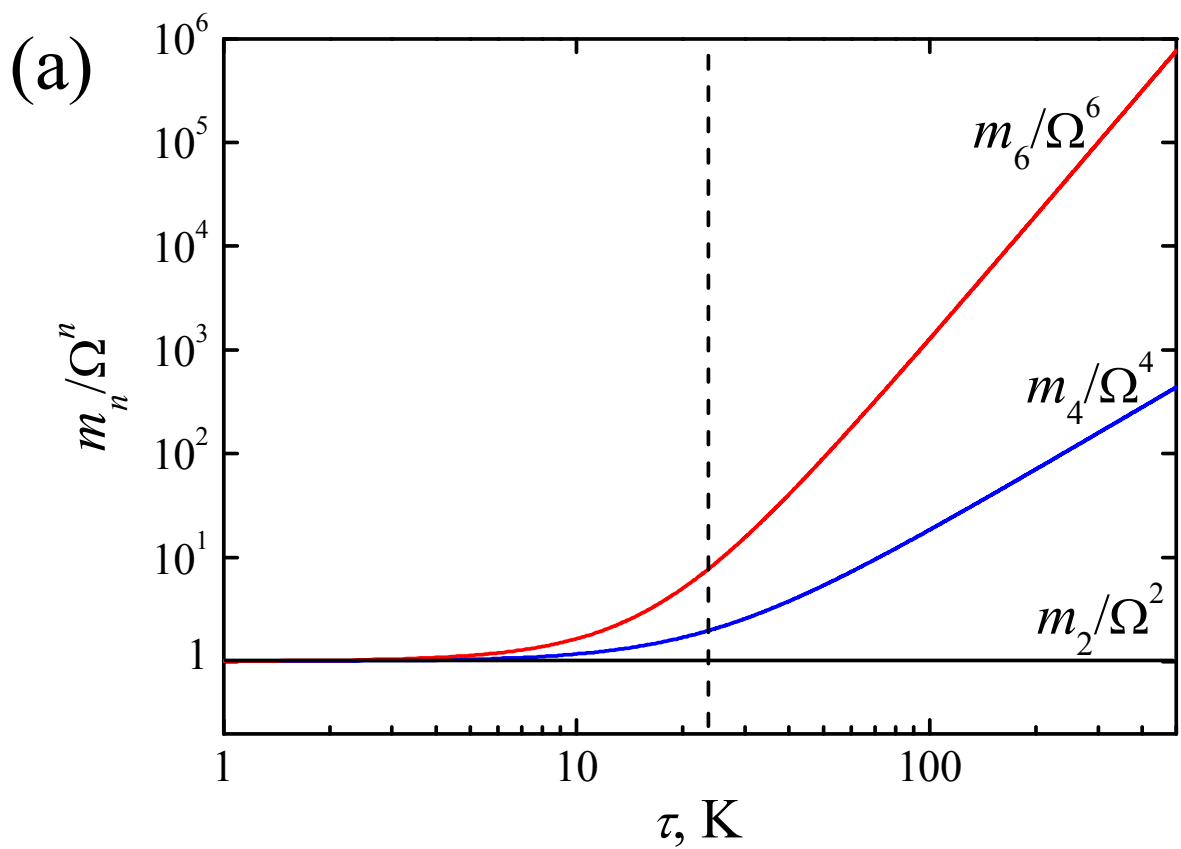

Fig. 2b

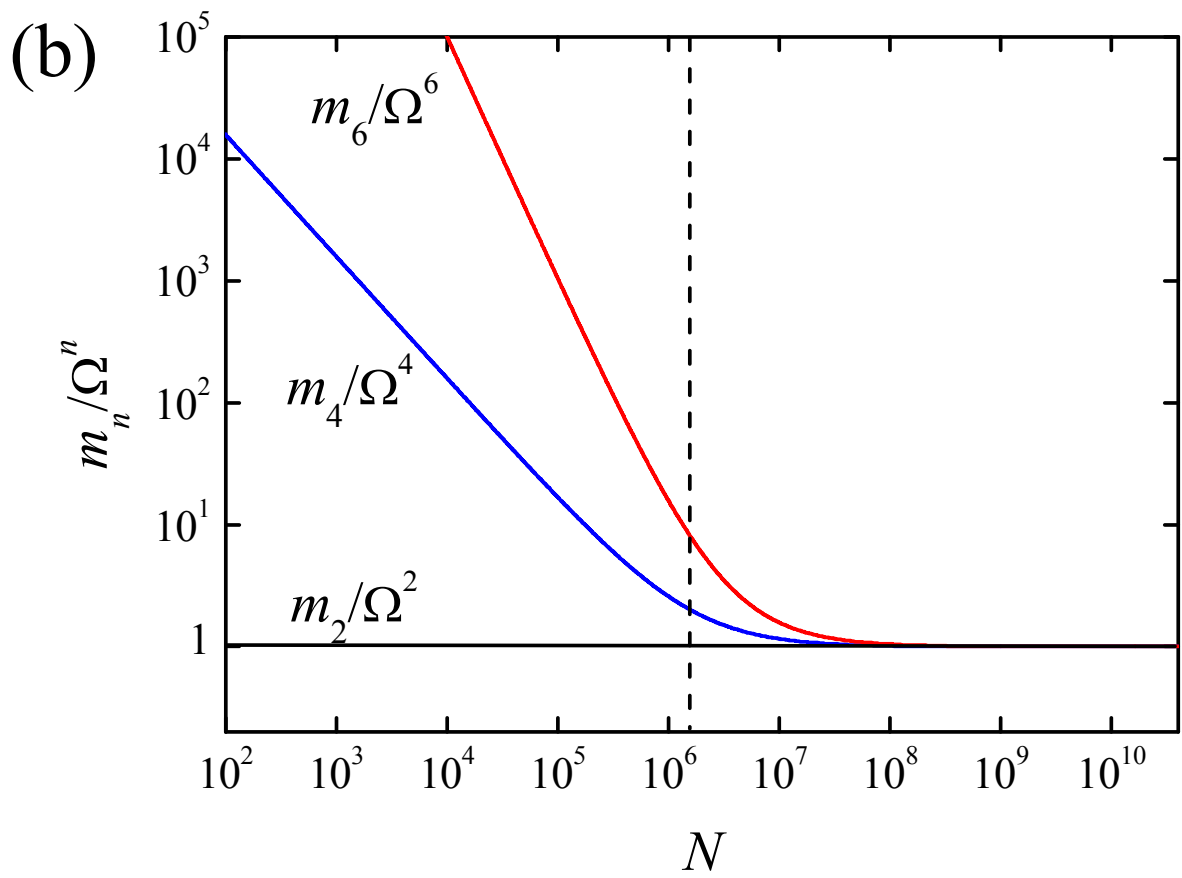

Fig. 3

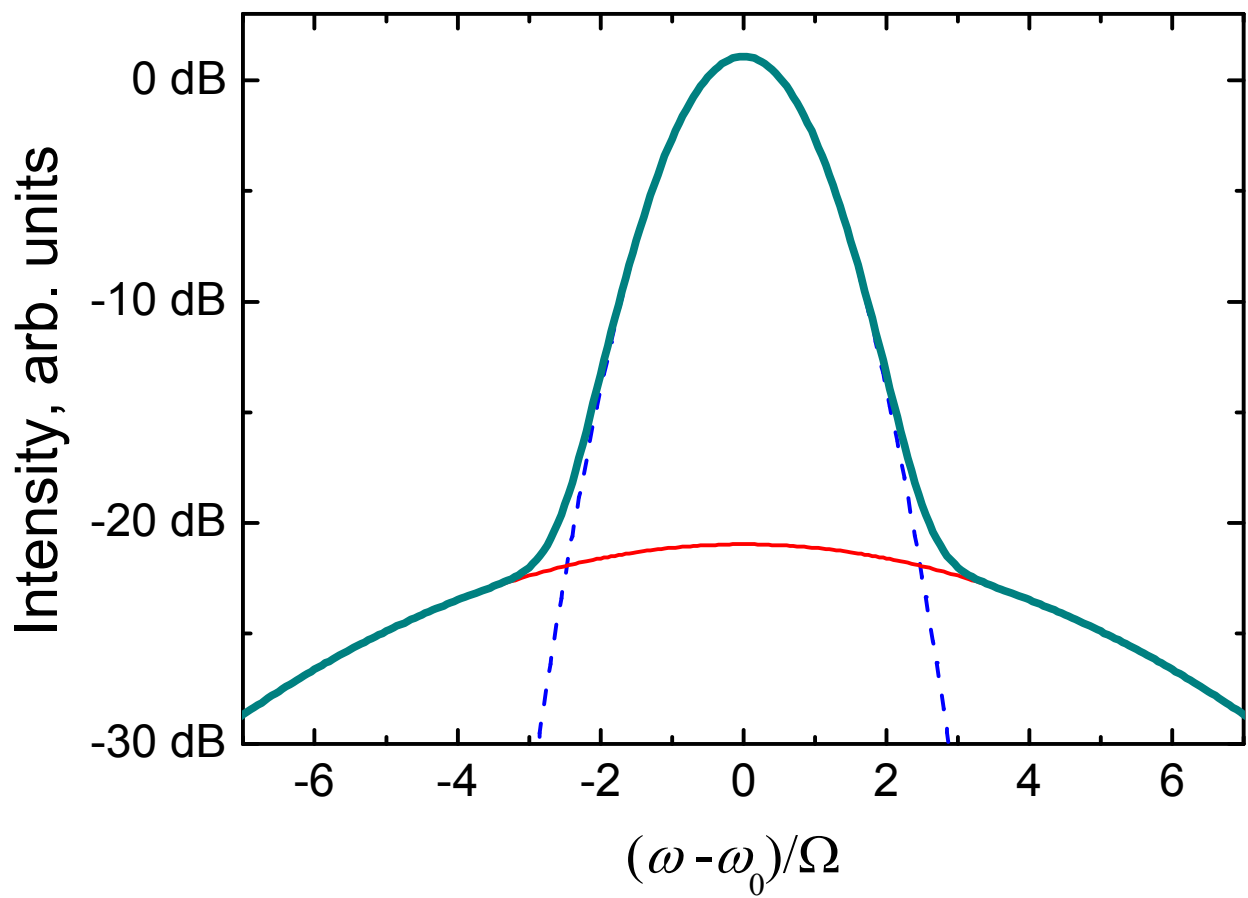

**Tables:**

Table 1. The emission spectrum of thermally equilibrated spin ensemble at resonance with the microwave cavity mode $(\omega_S = \omega_0)$ and the possibility of the strong spin-photon coupling. $\tau_C = \omega_S \sqrt{N}/2$ is the critical temperature of the system.

| Temperature interval | Emission spectrum | Realization of strong spin-photon coupling |
|---|---|---|
| $0 \leq \tau/\tau_C \ll N^{-1/2}$ (low temperatures) | Two delta functions separated by the gap $2g\sqrt{N}$. | possible |
| $N^{-1/2} \ll \tau/\tau_C \ll 1$ (medium temperatures) | Two Gaussian lines with equal standard deviations split by the gap $g\sqrt{2N\omega_S/\tau}$ | possible |
| $N^{-1/2} \ll \tau/\tau_C \ll \sqrt{N}$ (high temperatures) | Superposition of narrow and wide lines with half-widths $\sim g\sqrt{N\omega_S/\tau}$ and $\sim g\sqrt{\tau/\omega_S}$. Rabi splitting is unresolved. | hardly possible |
| $\tau/\tau_C > \sqrt{N}$ | Only the cavity mode is present. The photon field can be treated classically. | impossible |